\documentclass[twocolumn,showpacs,amsmath,amssymb,amsfonts,a4paper,aps,prd,10pt]{revtex4}
\usepackage{graphicx}
\usepackage{epstopdf}
\newcommand{\nc}{\newcommand}
\nc{\ba}{\begin{eqnarray}} \nc{\ea}{\end{eqnarray}}
\newcommand\be{\begin{equation}}
\newcommand\ee{\end{equation}}
\nc{\D}{\overline{\mbox{D3}}}

\nc{\ga}{\gamma} \nc{\tnu}{\tilde{\nu}} \nc{\tmu}{\tilde{\mu}}

\def\pp{^\prime}

\nc{\x}{{\bf{x}}}
\newcommand{\f}[2]{\frac{#1}{#2}}
\newcommand{\mc}[1]{\mathcal{#1}}

\begin{document}

\title{Massive cosmological scalar perturbations}
\author{Nima Khosravi$^{1,2}$}
\email{nima@aims.ac.za}
\author{Hamid Reza Sepangi$^{3}$}
\email{hr-sepangi@sbu.ac.ir}
\author{Shahab Shahidi$^{3}$}
\email{s_shahidi@sbu.ac.ir}

\affiliation{$^1$Cosmology Group, African Institute for Mathematical
Sciences, Muizenberg 7945, South Africa}\affiliation{$^2$South
African Astronomical Observatory, Observatory Road, Observatory,
Cape Town, 7935, South Africa} \affiliation{$^3$Department of
Physics, Shahid Beheshti University, G. C., Evin, Tehran 19839,
Iran}

\begin{abstract}
We study the cosmological perturbations of the new bi-metric gravity proposed by Hassan and Rosen \cite{hasan} as a representation of massive gravity. The mass term in the model, in addition of ensuring ghost freedom for both metrics, causes the two scale factors to mix at the cosmological level and this affects the cosmological perturbation of the model. We find two combinations corresponding to the entropy and adiabatic perturbations of the theory. In this sense we show that the adiabatic perturbations could be a source for the entropy perturbations. So in addition to the adiabatic perturbations, entropy perturbations can also be present in this theory. We also  show that the adiabatic perturbations are not constant at the super horizon scales, implying that the theory could not be used to describe the inflationary epoch, even if it can impose some corrections to the standard inflationary scenarios.

\end{abstract}
\maketitle

\section{Introduction}
The current problems facing standard cosmology are of such  breath and depth that any
fixation to these problems requires a deeper understanding of the prevailing theory that
has been and is still in use, namely the general theory of relativity.
Novel ideas, both old and new, have been suggested over its rather long history
in the hope of a partial remedy to some of the most pressing problems we are facing today.
Two interesting observations made over the period of the  past two decades, namely, the accelerated expansion of
the universe and galaxy rotation curves, amongst others, have stirred a plethora of activities
aimed at modifying the standard Einstein-Hilbert (EH) action and hence to offer a solution to some
of these problems. With partial success, one may consider various modifications to the standard EH action in the form
of modified theories of gravity with and without torsion, gravity in extra dimensions,
and more recently, massive gravity, to name but a few. The latter has been attracting the attention of many experts in the field in the past few years. It is based on an old idea by Fierz and Pauli (FP) \cite{fierz} where an effective field theory with a massive graviton was proposed. This theory
was, however, problematic at the linearized level since the corresponding Newtonian potential was discontinuous
for a vanishing mass, $m^2$, resulting, for example, in a large correction to the deflection of light around the Sun compared to the experimentally accepted value \cite{vd} predicted by General Relativity (GR). This is referred to as
the vDVZ discontinuity. The source of the discrepancy was traced to the degrees of freedom of the graviton,
being two for a massless and five for a massive graviton. Another setback was discovered later on when it was demonstrated
that when self interacting terms are added to the action, ghosts would appear in the theory \cite{boul}. It has
only been in the past few years that a method for fixing the above problems has been realized \cite{gab}. Theories
rooted in the FP action are  collectively known as massive gravity.

Perturbation methods are and have been an integral part of attempts
to find solutions to complicated problems. Massive gravity is
therefore no exception in this regard. In non-linear theories such
as GR, perturbation methods can be particularly useful when one
seeks the effects of small changes in the metric. Such methods have
been exploited in recent years to study, for example, the structure
formations in the universe. It then seems only natural to conduct
such a study when dealing with massive gravity. Such a study becomes
even more attractive if one considers the fact that massive gravity
is inherently a bi-metric theory. This comes about since any
modification to the EH action in the form of a self-coupling term
involving no derivative whose definition is based on one particular
metric requires an additional metric which may be dynamical or fixed
\cite{dam,salam}. The appearance of a second metric in the theory
can be understood on general grounds. If the second metric is
non-singular, spherically symmetric, and both are diagonal in some
coordinate system, a Killing horizon for one metric must be a
Killing horizon for the other \cite{def}. Interestingly, it has been
shown that the off diagonal elements undergo no modification at
large distances \cite{berz}. From a cosmological perspective, in a
bi-metric massive gravity theory with the second metric static,
there is no spatially flat FRW solutions \cite{gd}, contrary to the
bi-gravity formulation of the FRW cosmology for which homogeneous
solutions exist \cite{dcom,nnhs}.

It has increasingly been realized that modern observational date can be
explained by perturbation methods which have become an
integral  part of any study dealing with cosmological
fluctuations. On the other hand, from a theoretical point of view, the inflationary scenario can also be supported by the modern observational data rather accurately. This makes the studies
of the cosmological perturbations in massive gravity \cite{shinji} all the more interesting since
its predictions are gradually becoming known and are still
not on firm grounds. As was mentioned  above, one realization of massive gravity is in
the language of bi-metric theories. Consequently one may be
interested in a cosmological perturbation theory in such models, for
the existence of two metrics in these theories may result in
non-trivial features, specially in the behavior of relative metric
perturbations. In this paper we will consider scalar perturbations
for both metrics in a massive gravity model. We will obtain two gauge invariant combinations of the perturbed functions which can be responsible for the adiabatic and entropy perturbations. As a result we will obtain the super horizon limit of the equations of perturbations and show that the adiabatic perturbation is not constant in this scale.

The scope of the paper is as follows: in the next section we present the bi-metric model studied here. Section \ref{cosmo} deals with the the cosmological perturbations of the model and definition of the gauge invariant quantities. In section \ref{super} we obtain the equations of motion for the adiabatic perturbations and discuss the super horizon limit. Conclusions are drawn in section \ref{conclu}.

\section{The Model}\label{sec2}
We begin with the bimetric action \cite{hasan}
\begin{align}\label{eq200}
S=&-M_{g}^2\int d^4x\sqrt{-g}R(g)-M_{f}^2\int d^4x\sqrt{-f}R(f)\nonumber\\
&+2M^2_{eff}m^2\int d^4x\sqrt{-g}\sum^4_{n=0}\beta_n e_n\left(\sqrt{g^{-1}f}\right),
\end{align}
where $R(g)$ and $R(f)$ are the Ricci scalars corresponding to metrics $g_{\mu\nu}$ and $f_{\mu\nu}$ respectively, and $\beta_i$ are some arbitrary constants. We define the
polynomials $e_n$ as
\begin{align}\label{eq201}
&e_0(\mathbb{X})=1,\nonumber\\
&e_1(\mathbb{X})=[\mathbb{X}],\nonumber\\
&e_2(\mathbb{X})=\f{1}{2}\left([\mathbb{X}]^2-\left[\mathbb{X}^2\right]\right),\nonumber\\
&e_3(\mathbb{X})=\f{1}{6}\left([\mathbb{X}]^3-3[\mathbb{X}]\left[\mathbb{X}^2\right]+2\left[\mathbb{X}^3\right]\right),\nonumber\\
&e_4(\mathbb{X})=\textmd{det}\mathbb{X}.
\end{align}
with $\mathbb{X}=\sqrt{g^{-1} f}$. We also define
\begin{align}\label{eq202}
\f{1}{M_{eff}^2}=\f{1}{M_g^2}+\f{1}{M_f^2}.
\end{align}
Without the kinetic term for metric $f_{\mu\nu}$ the action describes a theory of massive gravity which is ghost free to all orders in the decoupling limit \cite{gab}. The cosmology of such a theory was studied in \cite{gd}.

The main point of the mass term is that it is symmetric in the metrics $f$ and $g$ in the sense that
\begin{align}\label{eq203}
\int d^4x&\sqrt{-g}\sum^4_{n=0}\beta_n e_n\left(\sqrt{g^{-1}f}\right)\nonumber\\
&=\int d^4x\sqrt{-f}\sum^4_{n=0}\beta_n e_{4-n}\left(\sqrt{f^{-1}g}\right).
\end{align}
So one can read off the equation of motion for $f_{\mu\nu}$ by having the equation of motion for $g_{\mu\nu}$.  The equation of motion for $g_{\mu\nu}$ can be easily obtained, resulting in \cite{hr}
\begin{align}\label{eq204}
M_g^2G_{\mu\nu}+m^2M_{eff}^2J_{\mu\nu}=0,
\end{align}
with
\begin{align}\label{eq205}
J_{\mu\nu}&=\beta_0g_{\mu\nu}-\beta_1\left(X_{\mu\nu}-g_{\mu\nu}e_1(\mathbb{X})\right)\nonumber\\
&+\beta_2\left(X^2_{\mu\nu}-X_{\mu\nu}e_1(\mathbb{X})+g_{\mu\nu}e_2(\mathbb{X})\right)\nonumber\\&-\beta_3\left(X^3_{\mu\nu}-X^2_{\mu\nu}e_1(\mathbb{X})+X_{\mu\nu}e_2(\mathbb{X})-g_{\mu\nu}e_3(\mathbb{X})\right),
\end{align}
where $G_{\mu\nu}$ is the Einstein tensor for metric $g_{\mu\nu}$.

The equation of motion for $f_{\mu\nu}$ is
\begin{align}\label{eq206}
M_f^2F_{\mu\nu}+m^2M_{eff}^2K_{\mu\nu}=0,
\end{align}
with
\begin{align}\label{eq207}
K_{\mu\nu}&=\alpha_0f_{\mu\nu}-\alpha_1\left(Y_{\mu\nu}-f_{\mu\nu}e_1(\mathbb{Y})\right)\nonumber\\
&+\alpha_2\left(Y^2_{\mu\nu}-Y_{\mu\nu}e_1(\mathbb{Y})+f_{\mu\nu}e_2(\mathbb{Y})\right)\nonumber\\
&-\alpha_3\left(Y^3_{\mu\nu}-Y^2_{\mu\nu}e_1(\mathbb{Y})+Y_{\mu\nu}e_2(\mathbb{Y})-f_{\mu\nu}e_3(\mathbb{Y})\right),
\end{align}
where $F_{\mu\nu}$ is the Einstein tensor for metric $f_{\mu\nu}$
and we have defined $\mathbb{Y}=\sqrt{f^{-1}g}$ and
$\alpha_n=\beta_{4-n}$. The cosmological solutions of this bi-metric
theory has been considered in \cite{dcom}.
\section{Cosmological Perturbations}\label{cosmo}
In order to study the cosmological perturbations of the model, we
consider the scalar metric perturbations of the two metrics (our
notation is compatible with that in \cite{mukhanov})
\begin{align}\label{eq208}
ds^2_g=a_1^2(t)\big[&-(1+2\phi_1)dt^2+2\partial_iB_1dx^idt\nonumber\\
&+[(1-2\psi_1)\delta_{ij}+2\partial_i\partial_jE_1]dx^idx^j\big],
\end{align}
and
\begin{align}\label{eq209}
ds^2_f=a_2^2(t)\big[&-(1+2\phi_2)dt^2+2\partial_iB_2dx^idt\nonumber\\
&+[(1-2\psi_2)\delta_{ij}+2\partial_i\partial_jE_2]dx^idx^j\big].
\end{align}
Let us define four zero order quantities for metrics $g_{\mu\nu}$
\begin{subequations}
 \begin{align}
X_1 & =\beta_1a_1^2a_2+2\beta_2a_1a_2^2+\beta_3a_2^3,\label{eq210.5a}\\
X_2 & =2\beta_0a_1^3+7\beta_1a_1^2a_2+8\beta_2a_1a_2^2+3\beta_3a_2^3,\label{eq210.5b}\\
X_3 & =\beta_0a_1^3+2\beta_1a_1^2a_2+\beta_2a_1a_2^2,\label{eq210.5c}\\
X_4 & =\beta_0a_1^3+3\beta_1a_1^2a_2+3\beta_2a_1a_2^2+\beta_3a_2^3.\label{eq210.5d}
\end{align}
\end{subequations}
and  $f_{\mu\nu}$
\begin{subequations}
 \begin{align}
Y_1 & =\beta_3a_2^2a_1+2\beta_2a_2a_1^2+\beta_1a_1^3,\label{eq211.5a}\\
Y_2 & =2\beta_4a_2^3+7\beta_3a_2^2a_1+8\beta_2a_2a_1^2+3\beta_1a_1^3,\label{eq211.5b}\\
Y_3 & =\beta_4a_2^3+2\beta_3a_2^2a_1+\beta_2a_2a_1^2,\label{eq211.5c}\\
Y_4 & =\beta_4a_2^3+3\beta_3a_2^2a_1+3\beta_2a_2a_1^2+\beta_1a_1^3.\label{eq211.5d}
\end{align}
\end{subequations}
These terms can simplify the mass parts of the equations of motion.
The background equations for metric $g_{\mu\nu}$ is
\begin{subequations}
 \begin{align}
&H_1^2=\f{1}{3}\f{m^2M_{eff}^2}{M_g^2}\f{1}{a_1}X_4,\label{eq210.6a}\\
&H_1^2+2H_1\pp=\f{m^2M_{eff}^2}{M_g^2}\f{1}{a_1}X_4\label{eq210.6b},
\end{align}
\end{subequations}
where $'\equiv d/dt$ and $H_1\equiv a_1'/a_1$. Similarly for $f_{\mu\nu}$ we have
\begin{subequations}
 \begin{align}
&H_2^2=\f{1}{3}\f{m^2M_{eff}^2}{M_f^2}\f{1}{a_2}Y_4,\label{eq210.7a}\\
&H_2^2+2H_2\pp=\f{m^2M_{eff}^2}{M_f^2}\f{1}{a_2}Y_4\label{eq210.7b}
\end{align}
\end{subequations}
where $H_2\equiv a_2'/a_2$. For the Kinetic term of the
metric $g_{\mu\nu}$, we obtain, to first order \footnote{Note that
in \cite{mukhanov} the Einstein tensor has been written with an
upper and a lower index but we write it with both lower indices.}
\begin{subequations}
\begin{align}
G^{(1)}_{00}&=2\nabla^2\bigg(\psi_1+H_1(E_1^\prime-B_1)\bigg)-6H_1\psi_1^\prime,\label{eq210a}\\
G^{(1)}_{0i}&=2\partial_i\bigg((\phi_1-\f{1}{2}B_1H_1)H_1-B_1H_1^\prime+\psi_1^\prime\bigg),\label{eq210b}\\
G^{(1)}_{ij}&=\partial_i\partial_j\bigg(\psi_1-\phi_1-B_1^\prime-4E_1H_1^\prime+2H_1E_1^\prime+E_1^{\prime\prime}\nonumber\\
&-2H_1(B_1+E_1H_1)\bigg),\label{eq210c}
\end{align}
and
\begin{align}
\sum_{i=1}^3G^{(1)}_{ii}&=6(\phi_1+\psi_1)
\left(H_1^2+2H_1^\prime\right)+
6H_1(2\psi_1^\prime+\phi_1^\prime)\nonumber\\
&+6\psi_1^{\prime\prime}+\nabla^2D,\label{eq210d}
\end{align}
\end{subequations}
with
\begin{align}\label{eq}
D=4H_1(B_1-E_1^\prime)&-2\left(H_1^2+2H_1^\prime\right)E_1\nonumber\\
&-2\left(\psi_1-\phi_1-B_1^\prime+E_1^{\prime\prime}\right).
\end{align}
The mass term then becomes, to the first order (see the Appendix)
\begin{subequations}
\begin{align}
J^{(1)}_{00}&=\f{1}{a_1}\bigg[X_1\bigg(3(\psi_2-\psi_1)+\nabla^2(E_1-E_2)\bigg)-2X_4\phi_1\bigg],\label{eq211a}\\
J^{(1)}_{0i}&=\f{1}{2a_1}\partial_i\bigg[X_2B_1-X_1B_2\bigg],\label{eq211b}\\
J^{(1)}_{ij}&=\f{1}{a_1}\partial_i\partial_j\bigg[X_2E_1-X_1E_2\bigg],\qquad i\neq j \label{eq211c}
\end{align}
and
\begin{align}
\sum_{i=1}^3J^{(1)}_{ii}&=\f{1}{a_1}\bigg[2X_3(\nabla^2E_1-3\psi_1)\nonumber\\
&+X_1(2\nabla^2E_2-6\psi_2+3\phi_2-3\phi_1)\bigg].\label{eq211d}
\end{align}
\end{subequations}
The kinetic term for the metric $f_{\mu\nu}$ is represented by its
corresponding Einstein tensor $F_{\mu\nu}$ which can be deduced for
the first order  perturbations by making the transformation
$1\rightarrow2$ in equations \eqref{eq210a}-\eqref{eq210d}. The mass
term can be obtained easily by  transformations
$1\leftrightarrow2$ and $\beta_n\rightarrow\beta_{4-n}$ in equations
\eqref{eq211a}-\eqref{eq211d}. The result is
\begin{subequations}
\begin{align}
K^{(1)}_{00}&=\f{1}{a_2}\bigg[Y_1\bigg(3(\psi_1-\psi_2)+\nabla^2(E_2-E_1)\bigg)-2Y_4\phi_2\bigg],\label{eq212a}\\
K^{(1)}_{0i}&=\f{1}{2a_2}\partial_i\bigg[Y_2B_2-Y_1B_1\bigg],\label{eq212b}\\
K^{(1)}_{ij}&=\f{1}{a_2}\partial_i\partial_j\bigg[Y_2E_2-Y_1E_1\bigg],\qquad i\neq j \label{eq212c}
\end{align}
and
\begin{align}
\sum_{i=1}^3K^{(1)}_{ii}&=\f{1}{a_2}\bigg[2Y_3(\nabla^2E_2-3\psi_2)\nonumber\\
&+Y_1(2\nabla^2E_1-6\psi_1+3\phi_1-3\phi_2)\bigg].\label{eq212d}
\end{align}
\end{subequations}
As we know, general relativity is invariant under coordinate
transformations \footnote{It is important to mention that a general
bi-metric model enjoys a diff$^2$-invariance. But as it is shown in
\cite{boulanger}, in the presence of an interaction term this
symmetry reduces to diff-invariance which is exactly the GR
symmetry.}. Hence we restrict our model to scalar perturbations so
we only consider scalar coordinate transformations as follow
\begin{subequations}
\begin{align}\label{eq212}
t&\rightarrow t+\delta t,\\
x^i&\rightarrow x^i+\delta^{ij}\partial_j\delta x,
\end{align}
\end{subequations}
and consequently the scalar metric perturbations behave like
\footnote{Note that we employ the scalar coordinate transformations
for both metrics. It implicitly means that we assume both metrics
live on the same manifold (this assumption may cause some
ambiguities in the definition of the physical distance and parallel
transportation which is out of the scope of the present paper.
However, as was mentioned in \cite{nnhs}, a natural way of resolving
this problem is by considering the average of both metrics as the
physical metric at least in the linear regime). However Arkani-Hamed
et al. \cite{arkani} have assumed that each metric belongs to a
certain manifold and the mass term effectively stitches these
manifolds together. In this point of view  each metric is invariant
under its corresponding coordinate transformation.}
\begin{subequations}
\begin{align}\label{eq213}
\phi_i&\rightarrow \phi_i-\f{1}{a_i}\left(a_i\delta t\right)\pp,\\
B_i&\rightarrow B_i+\delta t-\delta x\pp,\\
E_i&\rightarrow E_i-\delta x,\\
\psi_i&\rightarrow\psi_i+H_i\delta t,
\end{align}
\end{subequations}
where $i=1,2$. We then define six independent gauge invariant quantities as follows
\begin{subequations}
\begin{align}\label{eq214}
\Psi_i&=\psi_i+H_i(E_i\pp-B_i),\\
\Phi_i&=\phi_i-\f{1}{a_i}\left(a_i(E_i\pp-B_i)\right)\pp,\\
\mc{E}&=E_1-E_2,\label{eq214c}\\
\mc{B}&=B_1-B_2,\label{eq214d}
\end{align}
\end{subequations}
with $i=1,2$. Using  \eqref{eq214c} and \eqref{eq214d}, one can define a gauge invariant quantity
 \begin{align}\label{eq215}
\Lambda&=\mc{E}^\prime-\mc{B},
\end{align}
which simplifies the  calculations that follow.
In order to write the field equations in the gauge invariant form we
can fix two scalar gauge freedoms by the following conditions
\begin{align}\label{eq216}
E_1+E_2=0,\qquad B_1+B_2=0.
\end{align}
This gauge fixing has the advantage that the symmetry between the two metrics remains true, even after the gauge fixing.
We then rewrite the kinetic term to first order for  metric
$g_{\mu\nu}$ i.e. relations (\ref{eq210a})-(\ref{eq210d}) in terms
of six gauge invariant quantities $\Psi_{1,2}$, $\Phi_{1,2}$,
$\Lambda$ and $\mc{E}$ as
\begin{subequations}
\begin{align}
G^{(1)}_{00}&=2\nabla^2\Psi_1-3H_1(2\Psi_1-H_1\Lambda)\pp,\label{eq217a}\\
G^{(1)}_{0i}&=\partial_i\left[2(\Psi_1\pp+H_1\Phi_1)+\f{3}{2}H_1^2\Lambda-\f{1}{2}(H_1^2+2H_1\pp)\mc{E}\pp\right],\label{eq217b}\\
G^{(1)}_{ij}&=\partial_i\partial_j\left[\Psi_1-\Phi_1-(H_1^2+2H_1\pp)\mc{E}\right],\label{eq217c}
\end{align}
and
\begin{align}
\sum_{i=1}^3G^{(1)}_{ii}&=\nabla^2\left[2(\Psi_1-\Phi_1)-(H_1^2+2H_1\pp)\mc{E}\right]\nonumber\\
&+6\Psi_1^{\prime\prime}+6H_1(\Phi_1+2\Psi_1)\pp-3(H_1^{\prime\prime}+H_1H_1\pp)\Lambda\nonumber\\&+6(H_1^2+2H_1\pp)(\Phi_1+\Psi_1),\label{eq217d}
\end{align}
\end{subequations}
and similarly for  metric $f_{\mu\nu}$ we obtain \footnote{We note
that in order to read the equations of metric $f_{\mu\nu}$ from the
equations of  metric $g_{\mu\nu}$, one may make the transformations
$\Lambda\rightarrow -\Lambda$ and $\mc{E}\rightarrow -\mc{E}$.}
\begin{subequations}
\begin{align}
F^{(1)}_{00}&=2\nabla^2\Psi_2-3H_2(2\Psi_2+H_2\Lambda)\pp,\label{eq218a}\\
F^{(1)}_{0i}&=\partial_i\left[2(\Psi_2\pp+H_2\Phi_2)-\f{3}{2}H_2^2\Lambda+\f{1}{2}(H_2^2+2H_2\pp)\mc{E}\pp\right],\label{eq218b}\\
F^{(1)}_{ij}&=\partial_i\partial_j\left[\Psi_2-\Phi_2+(H_2^2+2H_2\pp)\mc{E}\right],\label{eq218c}
\end{align}
and
\begin{align}
\sum_{i=1}^3F^{(1)}_{ii}&=\nabla^2\left[2(\Psi_2-\Phi_2)+(H_2^2+2H_2\pp)\mc{E}\right]\nonumber\\
&+6\Psi_2^{\prime\prime}+6H_2(\Phi_2+2\Psi_2)\pp+3(H_2^{\prime\prime}+H_2H_2\pp)\Lambda\nonumber\\&+6(H_2^2+2H_2\pp)(\Phi_2+\Psi_2).\label{eq218d}
\end{align}
\end{subequations}
In terms of gauge invariant quantities, the mass terms
\eqref{eq211a}-\eqref{eq211d} can be rewritten as
\begin{subequations}
\begin{align}
J^{(1)}_{00}=\f{1}{a_1}\bigg[&\f{3}{2}X_1\big[2(\Psi_1+\Psi_2)+(H_2+H_1)\Lambda+\f{2}{3}\nabla^2\mc{E}\big]\nonumber\\&-X_4\big[2\Phi_1+H_1\Lambda+\Lambda\pp\big]\bigg],\label{eq219a}
\end{align}
\begin{align}
J^{(1)}_{0i}&=\f{X_1+X_2}{4a_1}\partial_i(\mc{E}\pp-\Lambda),\label{eq219b}\\
J^{(1)}_{ij}&=\f{X_1+X_2}{2a_1}\partial_i\partial_j\mc{E},\qquad i\neq j \label{eq219c}
\end{align}
\begin{align}
\sum_{i=1}^3J^{(1)}_{ii}&=\f{1}{a_1}\bigg[(X_3-X_1)\nabla^2\mc{E}
-6(X_3\Psi_1+X_1\Psi_2)\nonumber\\&+3(H_1X_3-H_2X_1)\Lambda\nonumber\\&+\f{3}{2}X_1\big[2(\Phi_2-\Phi_1)-(H_1+H_2)\Lambda-2\Lambda\pp\big]\bigg],\label{eq219d}
\end{align}
\end{subequations}
and similarly for \eqref{eq212a}-\eqref{eq212d}
\begin{subequations}
\begin{align}
K^{(1)}_{00}=\f{1}{a_2}\bigg[&\f{3}{2}Y_1\big[2(\Psi_2+\Psi_1)-(H_2+H_1)\Lambda-\f{2}{3}\nabla^2\mc{E}\big]\nonumber\\&-Y_4\big[2\Phi_2-H_2\Lambda-\Lambda\pp\big]\bigg],\label{eq220a}
\end{align}
\begin{align}
K^{(1)}_{0i}&=-\f{Y_1+Y_2}{4a_2}\partial_i(\mc{E}\pp-\Lambda),\label{eq220b}\\
K^{(1)}_{ij}&=-\f{Y_1+Y_2}{2a_2}\partial_i\partial_j\mc{E},\qquad i\neq j \label{eq220c}
\end{align}
\begin{align}
\sum_{i=1}^3K^{(1)}_{ii}&=\f{1}{a_2}\bigg[(Y_1-Y_3)\nabla^2\mc{E}-6(Y_3\Psi_2\nonumber\\&+Y_1\Psi_1)-3(H_2Y_3-H_1Y_1)\Lambda\nonumber\\&+\f{3}{2}Y_1\big[2(\Phi_1-\Phi_2)+(H_1+H_2)\Lambda+2\Lambda\pp\big]\bigg].\label{eq220d}
\end{align}
\end{subequations}
By transforming to a spatial Fourier space, $\nabla^2$ is replaced by
$-k^2$, where $k$ is the wave number. Using the above equations and after some
algebra we arrive at writing $\mc{E}$ and $\Lambda$ as functions of $\Phi_i$
and $\Psi_i$ and deduce the following constraints
\begin{align}\label{eqA11}
\mc{E}=A_1(\Psi_1-\Phi_1)=-A_2(\Psi_2-\Phi_2),
\end{align}
and
\begin{align}\label{eqL11}
\Lambda&=-C_1\bigg[2(H_1\Phi_1+\Psi_1\pp)-\f{1}{2A_1}\left(A_1(\Psi_1-\Phi_1)\right)\pp\bigg]
\nonumber\\&=C_2\bigg[2(H_2\Phi_2+\Psi_2\pp)-\f{1}{2A_2}\left(A_2(\Psi_2-\Phi_2)\right)\pp\bigg],
\end{align}
where
\begin{align}\label{eq}
\f{1}{A_1}&=H_1^2+2H_1\pp-m_g\f{X_1+X_2}{2a_1},\\
\f{1}{A_2}&=H_2^2+2H_2\pp-m_f\f{Y_1+Y_2}{2a_2},
\end{align}
\begin{align}\label{eq}
\f{1}{C_1}&=-m_g\f{X_1+X_2}{4a_1}+\f{3}{2}H_1^2,\\
\f{1}{C_2}&=-m_f\f{Y_1+Y_2}{4a_2}+\f{3}{2}H_2^2,
\end{align}
and
\begin{align}\label{eq}
m_g=\f{m^2M_{eff}^2}{M_g^2},\hspace{1cm} m_f=\f{m^2M_{eff}^2}{M_f^2}.
\end{align}
Using the background equations \eqref{eq210.6a}, \eqref{eq210.6b},
\eqref{eq210.7a} and \eqref{eq210.7b}, one can simplify the
quantities $A_i$ and $C_i$ as follows
\begin{align}\label{eqAC}
A_1&=\f{1}{2}C_1=\f{2a_1}{m_g}(2X_4-X_1-X_2)^{-1},\\
A_2&=\f{1}{2}C_2=\f{2a_2}{m_f}(2Y_4-Y_1-Y_2)^{-1},
\end{align}
The second constraint can be simplified with the use of \eqref{eqA11} and \eqref{eqAC} with the result
\begin{align}\label{eqA12}
-A_1(H_1\Phi_1+\Psi_1\pp)=A_2(H_2\Phi_2+\Psi_2\pp),
\end{align}
which can be used to simplify the calculations. The dynamical
equations then become
\begin{widetext}
\begin{align}\label{eqD1}
-2k^2\Psi_1-3H_1(2\Psi_1-H_1\Lambda)\pp=m_g\bigg[-\f{3}{2a_1}X_1\bigg(2(\Psi_2-\Psi_1)+(H_1+H_2)\Lambda-\f{2}{3}k^2\mc{E}\bigg)+\f{X_4}{a_1}\big(2\Phi_1+H_1\Lambda+\Lambda\pp\big)\bigg]
\end{align}
\begin{align}\label{eqD2}
-2k^2\Psi_2-3H_2(2\Psi_2+H_2\Lambda)\pp=m_f\bigg[-\f{3}{2a_2}Y_1\bigg(2(\Psi_1-\Psi_2)-(H_1+H_2)\Lambda+\f{2}{3}k^2\mc{E}\bigg)+\f{Y_4}{a_2}\big(2\Phi_2-H_2\Lambda-\Lambda\pp\big)\bigg]
\end{align}
and
\begin{align}\label{eqD3}
&-k^2\big[2(\Phi_1-\Psi_1)-(H_1^2+2H_1\pp)\mc{E}\big]+6H_1(\Phi_1+2\Psi_1)\pp+6\Psi_1^{\prime\prime}-3(H_1^{\prime\prime}+H_1H_1\pp)\Lambda+6(H_1^2+2H_1\pp)(\Phi_1+\Psi_1)\nonumber\\&=-\f{m_g}{a_1}\bigg[-k^2(X_3-X_1)\mc{E}-6(X_3\Psi_1+X_1\Psi_2)+3(H_1X_3-H_2X_1)\Lambda+\f{3}{2}X_1\big[2(\Phi_2-\Phi_1)-(H_1+H_2)\Lambda-2\Lambda\pp\big]\bigg]
\end{align}
\begin{align}\label{eqD4}
&-k^2\big[2(\Phi_2-\Psi_2)+(H_2^2+2H_2\pp)\mc{E}\big]
+6H_2(\Phi_2+2\Psi_2)\pp+6\Psi_2^{\prime\prime}
+3(H_2^{\prime\prime}+H_2H_2\pp)\Lambda+
6(H_2^2+2H_2\pp)(\Phi_2+\Psi_2)\nonumber\\&=
-\f{m_f}{a_2}\bigg[k^2(Y_3-Y_1)\mc{E}-6(Y_3\Psi_2+Y_1\Psi_1)-3(H_2Y_3-H_1Y_1)\Lambda+\f{3}{2}Y_1\big[2(\Phi_1-\Phi_2)+(H_1+H_2)\Lambda+2\Lambda\pp\big]\bigg].
\end{align}
\end{widetext}
where we keep in mind that the quantities $\Lambda$ and $\mc{E}$
must be replaced by \eqref{eqA11} and \eqref{eqL11}. In this sense
we obtain four equations for four dynamical variables.

\section{Equations of motion and the super horizon limit}\label{super}
From \cite{nnhs} we know that a non-trivial classical solution for
the scale factors can be deduced by assuming $a_1=a_2$. With this
assumption the solution will be homogeneous de-Sitter solution. In
this case the adiabatic direction which should be along the
homogeneous solution is clearly along the straight line, $a_1=a_2$
in the phase space. It means that the adiabatic fluctuations should
have the same contribution from both metrics (\ref{eq208}) and
(\ref{eq209}). Or in other words, one expects that a suitable
combination of the variables $\Psi=\Psi_1+\Psi_2$ and
$\Phi=\Phi_1+\Phi_2$ should be responsible for the adiabatic perturbations.
Similarly, a suitable combination of the variables
$\psi=\Psi_1-\Psi_2$ and $\phi=\Phi_1-\Phi_2$ which is orthogonal to
the adiabatic one should be responsible for the entropy perturbations.

At this stage a discussion on the relation of  curvature
perturbation to $\Psi_1$ and $\Psi_2$ would be in order. As is well
known \cite{mukhanov}, curvature perturbation is defined in terms of
matter and metric perturbations. However the definition in a bi-metric model needs some care.
As has been mentioned previously \cite{nnhs}, a
convenient physical quantity for coupling to matter is the average
of the two metrics. This claim is based on the fact that by
assuming a exchangeable role for both metrics,  the only combination which is
allowed at the linear level is the sum of the first order
perturbations in the tensor mode \cite{nnhs} or in the scalar mode which
is the purpose of the present work. Therefore, if we were to add matter to
our formulation, a convenient combination representing curvature
perturbation would be that constructed from $\psi_1+ \psi_2$. The
absence of matter in the present discussion means that a suitable
gauge invariant quantity constructed out of $\psi_1+ \psi_2$ would
be $\Psi_1+\Psi_2$, which is the same as the adiabatic perturbation
defined above.

To solve equations \eqref{eqD1}-\eqref{eqD4}, we assume $a_1=a_2=a$ and define constants $x_i$ and
$y_i$ as
\begin{align}\label{eq}
X_i=a^3x_i,\qquad Y_i=a^3y_i,
\end{align}
which, using equations \eqref{eq210.6a} and \eqref{eq210.7a}, imply that
\begin{align}\label{eq}
m_gx_4=m_fy_4.
\end{align}
In order to write the equations in terms of the new variables $\Psi,\Phi,\psi$ and $\phi$, we define
\begin{align}\label{eq}
\alpha&=\f{1}{2}\left(\f{x_4}{x_1}+\f{y_4}{y_1}\right),\\
\beta&=\f{1}{2}\left(\f{x_4}{x_1}-\f{y_4}{y_1}\right).
\end{align}
The constraint equations \eqref{eqA11} and \eqref{eqA12} can then be written as
\begin{align}\label{eqcond1}
&\alpha(\Psi-\Phi)+\beta(\psi-\phi)=0,\\\label{eqcond2}
&\alpha(H\Phi+\Psi\pp)+\beta(H\phi+\psi\pp)=0.
\end{align}
Equations \eqref{eqD1} and \eqref{eqD2} take the form
\begin{align}\label{eqD5}
6H\Psi\pp&+(2 k^2+\frac{\beta^2}{\alpha^2-\beta^2}k^2)\Psi\nonumber\\&+(6H^2-\frac{\beta^2}{\alpha^2-\beta^2}k^2)\Phi+\f{18\beta}{\alpha^2-\beta^2} H^3\Lambda\nonumber\\&=\frac{\beta \alpha}{\alpha^2-\beta^2}k^2(\phi-\psi)+\f{18\beta}{\alpha^2-\beta^2} H^2\psi,
\end{align}
and
\begin{align}\label{eqD6}
6H\psi\pp&+(2 k^2-\f{\alpha^2}{\alpha^2-\beta^2}k^2+\f{18\alpha}{\alpha^2-\beta^2} H^2)\psi\nonumber\\&+(\f{\alpha^2}{\alpha^2-\beta^2}k^2+6H^2)\phi-\f{18\alpha}{\alpha^2-\beta^2} H^3\Lambda\nonumber\\&=\f{\alpha\beta}{\alpha^2-\beta^2}k^2(\Psi-\Phi).
\end{align}
Also one can rewrite the relation \eqref{eqL11} as
\begin{align}\label{eq}
\Lambda&=\f{\alpha^2-\beta^2}{3\alpha H^2}\left[\f{3}{2}\psi\pp+\f{1}{2}\phi\pp+H\psi+H\phi\right]\nonumber\\
&=\f{\beta^2-\alpha^2}{3\beta H^2}\left[\f{3}{2}\Psi\pp+\f{1}{2}\Phi\pp+H\Psi+H\Phi\right].
\end{align}
Upon substituting $\Lambda$ we obtain
\begin{align}\label{eq52}
&\Psi'+\Phi'-\f{k^2}{3H}\Phi+\left(2H-\f{k^2}{3H}\right)\Psi=\f{-6\beta}{\alpha^2-\beta^2}H\psi,\\\label{eq53}
&\psi'+\phi'-\f{k^2}{3H}\phi+\left(2H-\f{6\alpha}{\alpha^2-\beta^2}H-\f{k^2}{3H}\right)\psi=0.
\end{align}
Equation \eqref{eq52} has an important property, showing that the
entropy perturbation can be a source for the adiabatic perturbation.

At this stage it is appropriate to take the following steps to make the equations less complicated and therefore easier to solve. From the first condition \eqref{eqcond1}
we have
\begin{align}\label{s1}
\alpha\Psi+\beta\psi=\alpha\Phi+\beta\phi,
\end{align}
and by the weighted addition of equations in \eqref{eq52} and
\eqref{eq53} one finds
\begin{align}\label{s2}
&\big(\alpha\Psi+\beta\psi)'+(\alpha\Phi+\beta\phi)'-\f{k^2}{3H}(\alpha\Phi+\beta\phi)\\&
+\left(2H-\f{k^2}{3H}\right)(\alpha\Psi+\beta\psi)=0.
\end{align}
Now by plugging \eqref{s1} into the above equation we have
\begin{align}\label{s3}
&\big(\alpha\Psi+\beta\psi)'
+\left(H-\f{k^2}{3H}\right)(\alpha\Psi+\beta\psi)=0.
\end{align}
The second condition \eqref{eqcond2} results in
\begin{align}\label{s4}
&\big(\alpha\Psi+\beta\psi)'
=-H\big(\alpha\Phi+\beta\phi),
\end{align}
and  reduces to
\begin{align}\label{s5}
&\big(\alpha\Psi+\beta\psi)'
+H\big(\alpha\Psi+\beta\psi)=0,
\end{align}
due to \eqref{s1}. Comparison of \eqref{s3} with \eqref{s5} shows
\begin{align}\label{s6}
\f{k^2}{3H}\big(\alpha\Psi+\beta\psi)=0,
\end{align}
which leads to
\begin{align}\label{s6.5}
\alpha\Psi=-\beta\psi,
\end{align}
and consequently from \eqref{s1}
\begin{align}\label{s7}
\alpha\Phi=-\beta\phi.
\end{align}
We note that \eqref{s6.5} and \eqref{s7} render equations \eqref{eq52} and
\eqref{eq53} equivalent. So by plugging the above
relations into \eqref{eq52} one finds
\begin{align}\label{s8}
&(\Psi+\Phi)'-\f{k^2}{3H}(\Psi+\Phi)+2\gamma H\Psi=0,
\end{align}
with
\begin{align}\label{eq}
\gamma=1-\f{3\alpha}{\alpha^2-\beta^2}.
\end{align}
The same procedure as above can be employed for the remaining
equations \eqref{eqD3} and \eqref{eqD4}. They, in conjunction with
equations \eqref{s6.5} and \eqref{s7}, can be used to arrive at an
equation of the form
\begin{align}\label{eqD5}
\Psi^{\prime\prime}+\Phi^{\prime\prime}+H(\Psi\pp+\Phi\pp)+2\gamma
H^2(2\Psi-\Phi)=0,
\end{align}
which is independent of \eqref{s8}. So for four variables $\psi$,
$\phi$, $\Psi$ and $\Phi$ we have four independent equations
\eqref{s6.5}, \eqref{s7}, \eqref{s8} and \eqref{eqD5}. The two latter
equations can be converted to a first order differential equation for
$\Psi$ and $\Phi$ by differentiating equation \eqref{s8} and
subtract the result from \eqref{eqD5}. The result is
\begin{align}\label{s9}
&\left(\f{k^2}{3H}+H\right)(\Psi+\Phi)\pp-\left(\f{k^2}{3}+2\gamma H^2\right)\left(\Psi+\Phi\right)\nonumber\\&-2\gamma H\Psi\pp
+4\gamma H^2\Psi=0.
\end{align}
Equations \eqref{s8} and \eqref{s9} constitute our final resulting
equations for $\Psi$ and $\Phi$. Now, combining the above equation
with \eqref{s8} results in
\begin{align}\label{s10}
(\Psi+\Phi)''-2H\left(\Psi+\Phi\right)'+\left(\f{k^2}{3}-2\gamma
H^2\right)(\Psi+\Phi)=0.
\end{align}
Since $H=-1/t$ as a consequence of the background equations, the
solution of the above equation is
\begin{align}\label{s11}
\Psi+\Phi=c_1 j_{n}\left(\frac{k t}{\sqrt{3}}\right)+c_2
y_{n}\left(\frac{k t}{\sqrt{3}}\right)
\end{align}
where $c_1$ and $c_2$ are integration constants, $j_n(x)$ and
$y_n(x)$ are spherical Bessel functions and
$$
n=\frac{-1+\sqrt{1+8\gamma}}{2}.
$$
Consequently, using the above
solutions and either of equations \eqref{s8} or \eqref{s9}, one can find
$\Psi(t)$ and $\Phi(t)$ separately. This means that by employing
equations \eqref{s6.5} and \eqref{s7}, the functions $\psi(t)$ and $\phi(t)$ and
eventually $\psi_1$, $\psi_2$, $\phi_1$ and $\phi_2$ can be found as functions of time.

For super-horizon modes, i.e. $k\sim0$,  equation \eqref{s10}
results in
\begin{align}\label{s11-2}
\Psi+\Phi=c_1 t+c_2t^{-2},
\end{align}
and by considering the fact that $t$ is the conformal time and
$t\in\{-\infty,0\}$, the  term ``$c_1 t$" can be considered as the damping term. The
growing mode, i.e. ``$c_2 t^{-2}$" term, results in
\begin{align}\label{s14}
\Psi=-c_2\f{1}{\gamma}t^{-2},\hspace{1cm}\Phi=c_2\f{1+\gamma}{\gamma}t^{-2}
\end{align}
This result suggests that in the theory of bi-metric gravity that we have
studied in this paper, the curvature perturbation is not constant at
the super horizon limit. This result is compatible with
\cite{bigravinf}. However in \cite{bigravinf} the authors have
studied an old bi-metric model in the presence of the cosmological
constant and the Fierz-Pauli mass term. In contrast,
we have studied a ghost free bi-metric model without any cosmological
constant and with a non-linear generalization of the Fierz-Pauli mass
term. The mass term in such a theory cannot then be used as a manifestation
of inflation. At the background level there is no way to a graceful exit
from inflation. In order to achieve such an exit, one should not consider
the mass as a constant, rather, it should be considered as a function of
some other fields \cite{gd}. Now, if one assumes the mass to be a function
of the inflaton field then the possibility of having a reduced mass at the
end of inflation would not be far-fetched and therefore a graceful exit from
de-Sitter phase would become a reality \cite{future}.

On the other hand it seems that the observational
data, e.g. the cosmic microwave background radiation, are consistent
with an inflationary scenario with nearly adiabatic and Gaussian
fluctuations. So the results of this paper are not compatible with
observation except by imposing some constraints on the entropy
mode's effects. Therefore, the consideration of a constant mass and the addition of some sort of
an inflationary scenario to the theory should result
in imposing some corrections to the curvature perturbations at the
super horizon limit which restrict the parameters of the theory to
that of the slow roll parameters. Eventually we expect that the
observational data constrains the mass of the graviton in this scenario.
However, this result may be acceptable because massive gravity is
responsible for the late time accelerated expansion of the universe
\cite{parisi}.

\section{Conclusions}\label{conclu}
Modern observational data have become so accurate that meaningful
comparisons with theoretical predictions are now a reality. This
has led to the introduction of various calculational techniques,
including metric perturbation methods,  which have been playing an
increasingly important role. In this paper we have employed such a method to
study a bi-metric massive gravity theory. The present work proposes
a full discussion of the scalar perturbations of the bi-metric
gravity, with all the potential terms at our disposal. Because the
mass term mixes the scale factors of the two metrics, we expect both
the  entropy and adiabatic perturbations to be non-zero and depend
upon each other. This can also be seen from the dynamical equations
of the model. As a result, similar to the double scalar field models
\cite{wands}, one may expect that the adiabatic perturbation should
be along the path of the classical solutions in the phase-space and the
entropy perturbation to be orthogonal to it. According to \cite{nnhs}, a
classical solution in the framework used here is $a_1=a_2$. So it seems
natural to think that the sum of two scalar perturbations
corresponds to the adiabatic perturbation and the entropy perturbation
corresponds to the difference of the two scalar fields. Since we
have four scalar perturbations, the suitable combinations are the
terms $\Psi_1\pm\Psi_2$ and $\Phi_1\pm\Phi_2$.  From equation
\eqref{eq52}, one can see explicitly that the entropy perturbations
are a source of the adiabatic perturbations which is in agreement with
double scalar field models \cite{wands}.

Interestingly, equation \eqref{s11-2} shows that the super horizon
adiabatic perturbations are not constant for the bi-metric model
discussed in this paper. We may expect that the mass term for
graviton can only impose some corrections to the results of the
inflationary scenario which we must add to the theory using other
methods. In this scenario, the mass term in the model becomes
restricted by the slow roll parameters. This can be the subject of
a future work.
\begin{center}
 \textbf{Note added}
\end{center}
During the completion of the present work, a study of the same subject appeared in \cite{come}.

\begin{acknowledgments}
The authors would like to thank Hassan Firouzjahi and Zahra Haghani
for useful comments. We are also grateful to the anonymous referee
for constructive comments which have improved the quality of the paper.
\end{acknowledgments}
\section{appendix}\label{ap}
In this appendix we derive the first order approximation to
matrix $\mathbb{X}=\sqrt{g^{-1}f}$.  Using equations \eqref{eq208}
and \eqref{eq209} we obtain, to first order in perturbation, for
matrix $\mathbb{G}=g^{-1}f$
\begin{align}\label{eq}
\mathbb{G}=\mathbb{G}^{(0)}+\epsilon\mathbb{G}^{(1)}+\mathcal{O}(\epsilon^2),
\end{align}
where
\begin{align}\label{eq}
\left(\mathbb{G}^{(0)}\right)^\mu_\nu=\left(\f{a_2}{a_1}\right)^2\delta^\mu_\nu,
\end{align}
and
\begin{align}\label{eq}
\left(\mathbb{G}^{(1)}\right)^0_0&=-2\left(\f{a_2}{a_1}\right)^2(\phi_1-\phi_2),\\
\left(\mathbb{G}^{(1)}\right)^0_i&=\left(\f{a_2}{a_1}\right)^2\partial_i(B_1-B_2),\\
\left(\mathbb{G}^{(1)}\right)^i_j&=-2\left(\f{a_2}{a_1}\right)^2[(\psi_1-\psi_2)\delta^i_j+\partial_i\partial_j(E_1-E_2)].
\end{align}
We finally expand  the matrix $\mathbb{X}=\sqrt{\mathbb{G}}$
\begin{align}\label{eq}
\mathbb{X}=\mathbb{X}^{(0)}+\epsilon\mathbb{X}^{(1)}+\mathcal{O}(\epsilon^2),
\end{align}
where
\begin{align}\label{eq}
\left(\mathbb{X}^{(0)}\right)^\mu_\nu=\f{a_2}{a_1}\delta^\mu_\nu,
\end{align}
and
\begin{align}\label{eq}
\left(\mathbb{X}^{(1)}\right)^\mu_\nu=\f{a_2}{2a_1}\left(\mathbb{G}^{(1)}\right)^\mu_\nu.
\end{align}
The expansion of  matrix $\mathbb{Y}=\sqrt{f^{-1}g}$ can be easily obtained by observing that metrics $g_{\mu\nu}$ and $f_{\mu\nu}$ can be transformed to each other by the transformation $1\rightarrow2$.


\end{document}